\documentclass[%
 reprint,
superscriptaddress,
nofootinbib,
 amsmath,amssymb,
 aps,
]{revtex4-2}

\usepackage{graphicx}
\usepackage{dcolumn}
\usepackage{bm}
\usepackage{amssymb}
\usepackage{amsmath}
\usepackage{amsfonts}
\usepackage{graphicx}
\usepackage{color}
\usepackage{xspace}
\usepackage[normalem]{ulem}
\usepackage{mathtools}
\usepackage{hhline}
\usepackage{braket}
\usepackage{subcaption}
\usepackage{float}
\usepackage[dvipsnames]{xcolor}
\usepackage{url}
\usepackage{orcidlink}

\begin{document}


\title{Measuring Coherent Radio and Microwave Photons from the Solar Corona}

        \author{Liang Chen\orcidlink{0000-0002-0224-7598}}
        \email{bqipd@pm.me}
        \affiliation{School of Fundamental Physics and Mathematical Sciences,
Hangzhou Institute for Advanced Study, UCAS, Hangzhou 310024 ,China}
        \affiliation{University of Chinese Academy of Sciences, 100190 Beijing, China}

                \author{Zizang Qiu
                \orcidlink{0009-0009-5338-4793}}
     \email{zizang.qiu@ed.ac.uk}
         \affiliation{School of Physics and Astronomy, University of Edinburgh, Edinburgh, EH9 3FD, United Kingdom}

        \author{Thomas W. Kephart\orcidlink{0000-0001-6414-9590}}        \email{tom.kephart@gmail.com}
        \affiliation{Department of Physics and Astronomy, Vanderbilt University, Nashville, TN 37235, USA}

        \author{Arjun Berera\orcidlink{0000-0001-5005-7812}}
        \email{ab@ed.ac.uk}
        \affiliation{School of Physics and Astronomy, University of Edinburgh, Edinburgh, EH9 3FD, United Kingdom}

\begin{abstract}
The rates of production of radio/microwave N-identical  photons states $\ket N$ from stimulated emission in the solar atmosphere are estimated. Effects of various decohering factors are shown to be small.
Ground based measurements of these quantum states via the inverse HOM effect are proposed. We argue that a signal is detectable and far above the noise in several cases.

\end{abstract}

\maketitle


{\it Introduction} -- Photons in certain
frequency bands can sustain quantum coherence at interstellar
distances \cite{Berera:2020rpl}, which introduces quantum mechanical interests into astronomical observation.
By quantum coherence we mean N photons in some specified quantum state, which particularly in
this paper means they are all identical, with the same momentum and polarization.
On a much longer scale, quantum coherence at cosmological distances can also be preserved as decoherence effects from interactions with the cosmological medium and the expansion of the Universe are negligible \cite{Berera:2021xqa}.
Viability of using photons in X-ray, optical and microwave bands as candidates to establish interstellar quantum communications was studied in  Ref.~\cite{Berera:2022nzs} and more recently in Ref.~\cite{Boyle:2024obk}.

Stimulated emission of N identical photons produced
in a specified quantum state, which we denote as
$|N\rangle$, is a possible type of quantum coherent photonic signals.
We expect such signals may even come directly from the Sun, although
deep below the surface of the star, they quickly decohere through interactions. But stimulated emission in the stellar atmosphere could still yield surviving coherent signals with the quantum nature intact.
We proposed this possibility and did some initial estimates of the rates of potentially detectable $\ket 2$ states on a 1 m$^2$ area near Earth from stimulated emission occurring in the Solar corona, for characteristic optical, UV and X-ray emission lines \cite{Berera:2023fhd}.

Valuable discussions of the mechanism of stimulated emission can be found in Refs.~\cite{townsend_2012,svelto_2010}.
According to Fermi's Golden Rule, 
a Bose enhancement factor arises for the transition probability amplitude of stimulated photon emission,
leading to a higher chance of obtaining states with a large number of photons of the same momentum and polarization. 
In Ref.~\cite{Berera:2023fhd}, we re-expressed the cross section of stimulated emission in terms of the wavelength $\lambda_0$, refraction index $n$ of the material, wavelength 
width of the emission $\Delta\lambda$, and rate of spontaneous emission $A$, as
\begin{equation}  \label{cross-section-se0}
\sigma_0     =   \left({\lambda_0\over2\pi n}\right)^2   \left( { A \lambda_0 \over c } \right) \left(  { \lambda_0 \over\Delta\lambda}\right) ~.
\end{equation}
Using  NIST data \cite{NIST} for the  numerical values of $A$ allows us to calculate cross sections of any stimulated emission line from any atom or ion. 

A single photon state $\ket 1$ traveling through a layer of type $a$ excited atoms with number density $n_a$  can become a multi-photon state $\ket N$ after a distance  $x$ due to stimulated emission. Here the distance $x$ is not used to localize the photon but only to label the location where measurements may be performed.
Denoting the probability of finding an $\ket N$ state by $P(x, N)$, Ref. \cite{Berera:2023fhd} shows that 
\begin{equation}   \label{N-probability}
P(x, N) = e^{-n_a\sigma_0 x} \left( 1 - e^{-n_a\sigma_0 x} \right)^{N-1}\;.
\end{equation}
  $P(x,N)$  is normalized to $1$, when summed over all $N$, independent of $x$.

We have done an extensive search of possible stimulated emission
processes in the solar atmosphere. In this Letter we report identification
of a process in the microwave region that we will show produces a sizable
rate and at the same time is amenable to measurement by a ground based
experiment on Earth.  The Letter starts by first providing a broader
range of stimulated emission processes giving estimates 
for how uncertainties are treated.
Next we identify potential sources of decoherence and show that these are
not significant.  We then discuss the uncertainties involved and compare the
rates for the microwave to radio band to those of Optical, UV, X-ray given
in Ref.~\cite{Berera:2023fhd}. We also propose how measurements could be
performed to detect these quantum states based on the HOM
effect\cite{Hong:1987zz}, focusing in
particular on the microwave range, which is the key case
of interest in this Letter.

\medskip

{\it Rates estimations} -- We estimate the rate of $\ket 1$-states produced by the Sun based on the measured values of the photon fluxes for various frequencies. As these $\ket 1$ states travel through the solar corona, they can stimulate a layer of excited atoms and become $\ket 2$ or $\ket 3$ states and so forth. We then determine the rate of these $\ket N$ states being detectable on a $1\ {\rm m}^2$ area near the Earth. 

We shall first consider photons in the radio to microwave regions.
Thousands of solar radio bursts occurring on different dates in the frequency range of 100-500 MHz observed by Ikarus/Zurich were analyzed by Aschwanden et al. \cite{1994ApJ...431..432A}. In this frequency range, we select 117 MHz to investigate since it is a transition line between two energy levels of hydrogen. The minimum solar flux between 105 MHz and 120 MHz was measured to be around 140 solar flux unit(SFU) \cite{1994ApJ...431..432A}, which equals $1.4\times10^{-20}$ W/(m$^2\cdot$Hz) and translates to an energy flux of $2.1\times10^{-13}$ W/m$^2$.

The ratio $\Delta\lambda/\lambda_0=0.1\%$ is a
reasonable estimate ~\cite{svelto_2010}
to use in eq.\eqref{cross-section-se0} to account for the stimulated emission linewidth. This implies an energy flux of $1.5\times10^{-15}$ W/m$^2$ and a photon number flux of $1.94\times10^{10}$ s$^{-1}$ m$^{-2}$ at 117 MHz. Thus we take the rate of $\ket 1$ state as $N_1(117\text{ MHz})\sim1.94\times10^{10}$.

The 117 MHz comes from transition between two energy levels of neutral hydrogen. But hydrogen will be ionized in deep corona regions because of the high temperature. Therefore, the stimulated emission could only occur across a thin layer just above photosphere. The thickness of the layer is about $L=2\times10^6$ m according to the electron density and temperature model of the chromosphere \cite{1990ApJ...355..700F, doi:10.1098/rsta.1976.0031}. 
We can think of the hydrogen atoms in this layer as being in thermal equilibrium since they are at the boundary of the photosphere. 
However, only those neutral hydrogen atoms that are in the upper level of the 117 MHz line are able to be stimulated by the 117 MHz photons. 
After calculating the Boltzmann factors of all the energy levels of a hydrogen atom listed by  NIST\cite{NIST} at temperature $T=5800$ K, the fraction of these stimulable hydrogen atoms is given by
\begin{equation}  \label{percent-stimulable}
{ e^{ - E_0 / k_B T} \over \sum_i e^{ - E_i / k_B T}  } \sim 4.53\times10^{-12}~,
\end{equation}
where $E_0$ is the energy of the stimulable hydrogen atoms, $E_i$ are the energies of hydrogen atoms at various energy levels and $k_B$ is the Boltzmann constant.
For the density of hydrogen atoms we take the modest
estimate of $10^{18}$ m$^{-3}$, though even higher
number densities are suggested, depending on the height above photosphere \cite{1990ApJ...355..700F, doi:10.1098/rsta.1976.0031}. Thus the density of stimulable hydrogen atoms is roughly $n_\text{H}=4.5\times10^{6}$~m$^{-3}$.

Substituting the rate $A=4.6551\times10^{-11}$ s$^{-1}$ \cite{NIST} of spontaneous emission of a 117 MHz photon from a hydrogen atom on the corresponding energy level into eq.\eqref{cross-section-se0}, results in a cross section $\sigma_0=6.6\times10^{-17}$ m$^2$ for stimulated emission. Given the density of hydrogen atoms and the cross section, the mean free path (MFP) of the stimulated emission is calculated to be about $3.34\times10^{9}$~m.

Following the path of a $\ket 1$ state, the probability of finding a $\ket 2$ state at a distance $L$ is $P(L, 2)$. Since we have estimated the rate of $\ket 1$-states $N_1(117\text{ MHz})$ in the discussion above, the rate of detectable $\ket 2$-states $N_2(117\text{ MHz})$ can be computed by applying eq. \eqref{N-probability},
\begin{flalign} \nonumber
N_2(117\text{ MHz}) =& N_1(117\text{ MHz}) \times e^{-n_\text{H}\sigma_0 L} \left( 1 - e^{-n_\text{H}\sigma_0 L} \right)
\\ \nonumber
\approx& 1.16\times10^7 ~\text{per s} ~.
\end{flalign}
This means that among all the 117 MHz photons impinging upon an area of 1 m$^2$ near Earth in one hour, there are over 11 million $\ket 2$ states.

Besides the 117 MHz band in the radio, we also find the rate of $\ket 1$-states for 232 MHz based on the time profile of radio flux from the Ikarus/Zurich observations \cite{1995ApJ...455..347A}. While in the microwave range, we gather the data of $\ket 1$-state rates for 1.078 GHz, 1.2378 GHz, 1.3711 GHz, 2.9334 GHz, 3.245 GHz and 9.9112 GHz respectively, from a spectral fit of the analysis of solar flares detected by the Owens Valley Solar Array(OVSA)\cite{Nita_2004}. These frequencies are in the radio to microwave range that could come from stimulated emission of hydrogen atoms \cite{NIST}. We did not find relevant rates of stimulated emission in this range from other atoms/ions. Following similar procedures laid out above, we present the calculated rates $N_2(\nu_0)$ of $\ket 2$-state for these frequencies $\nu_0$ in  table \ref{table1} below.
\begin{table}[ht]
\begin{center}
\begin{tabular}{  c|c|c|c|c } 
$\nu_0$ (GHz) & $N_1(\nu_0)$ s$^{-1}$  &  $N_2(\nu_0)$ s$^{-1}$ &  $N_3(\nu_0)$ s$^{-1}$ & $N_4(\nu_0)$ s$^{-1}$ \\\hline
0.117 & 1.94$\times10^{10}$  & $1.16\times10^7$ & 6953 & 4.16\\
0.232 & 1.20$\times10^{10}$  & $8.40\times10^6$ & 6093 & 4.41\\
1.078 & 1.30$\times10^{11}$  & $8.17\times10^6$ & 514 & 0.03\\
1.2378 & 9.85$\times10^{10}$  & $4.28\times10^7$ & 18688 &  8.14\\
1.3711 & 8.03$\times10^{10}$  & $9.31\times10^6$ & 1081 & 0.12\\
2.9334 & 1.75$\times10^{10}$  & $7.63\times10^6$ & 3327 & 1.44\\
3.245 & 1.43$\times10^{10}$  & $6.24\times10^6$ & 2719 & 1.18\\
9.9112 & 1.54$\times10^9$  & $2.87\times10^7$ & 547938 & 10447\\
\end{tabular}
\caption{\label{table1}
Estimates of the rates of $\ket 2$-state, $\ket 3$-state and $\ket 4$-state in the radio and microwave range.}
\end{center}
\end{table}
Across the spectrum, among the photons impinging upon an area of 1 m$^2$ near Earth in per second, millions of $\ket 2$-states, and from hundreds to half a million $\ket 3$-states could be measurable, depending on the efficiency of the instruments.
Even if the original measurements were performed by different research teams with different equipment at different locations and times, values of $N_2(\nu_0)$ in table \ref{table1} are of similar order of magnitudes.

Note that the hydrogen density varies by more than 9 orders of
magnitude in different parts of the solar atmosphere(for instance, the lowest particle density found in corona holes is $5\times10^{9}$ m$^{-3}$\cite{Morgan_2020}).
Thus it should be noted that the lower
value of this density then gives a cautious estimate for SE rates
which could be reduced by orders of magnitude comparing to those found in table \ref{table1}.  

Note that differences between values of $\ket 3$-state rates $N_3(\nu_0)$  in table \ref{table1} diverge, and the values of $\ket 4$-state rates $N_4(\nu_0)$ diverge further, varying from $3\times10^{-2}$ to $1\times10^4$. This divergence occurs as a manifestation of different MFPs for photons of different frequencies. For example, the MFPs for photons of 1.078 GHz and 9.9112 GHz are around $3.17\times10^{10}$ m and $1.04\times10^8$ m, respectively. To create a $\ket 4$-state from a $\ket 1$-state, there needs to be three consecutive events of stimulated emission. The two orders of magnitude difference in MFPs get cubed when calculating these rates, resulting in roughly six orders of magnitude difference between $N_4(1.078\text{ GHz})$ and $N_4(9.9112\text{ GHz})$.

Besides $\ket 3$-states and $\ket 4$-states, even higher number states can be produced for certain frequencies, which we show in the table below, on an adjusted time scale (per s, per hr, etc.).
\begin{table}[ht]
\begin{center}
\begin{tabular}{  c|c|c|c|c } 
$\nu_0$ (GHz) &  $N_5(\nu_0)$ &  $N_6(\nu_0)$ & $N_7(\nu_0)$ & $N_8(\nu_0)$  \\\hline
0.117 & 12 per hr  & &  \\
0.232 & 11 per hr    & &   \\
1.078 &    & &   \\
1.2378 &  12 per hr   & & \\
1.3711 &   &   &   \\
2.9334 & 2 per hr &  &   \\
3.245 & 1 per hr &   &  \\
9.9112 & 199 per s & 3 per s  & 260 per hr & 4 per hr 
\end{tabular}
\caption{\label{table2}
Estimates of the rates of $\ket 5$-state, $\ket 6$-state, $\ket 7$-state, and $\ket 8$-state in the radio and microwave range. If any rate is too low, we leave it blank. }
\end{center}
\end{table}
On an hourly basis, it is anticipated to have $\ket 5$-states for some of the frequencies, and for 9.9112 GHz, $\ket N$ up to $\ket 8$-states are expected.

In Ref.~\cite{Berera:2023fhd}, we performed estimates on the rates of $\ket 2$-states at 530.3 nm, 19.664 nm and 1.5 nm transition lines from the solar corona by picking the lower bound whenever a quantity is given in a range. Although being cautious and conservative, this approach underestimated the rates and we revisit the calculations here with more practical values of
quantities and so expect higher rates. 
For the 530.3 nm line,
we can assign a value of 20 CI for the intensity according to the observation data during the solar activity cycles \cite{165_Rusin},
which corresponds to a rate of $\ket 1$-states of $N_1^{530.3}\sim2.4\times10^{13}$. Also, we choose a value of  $1.416\times10^{4}$ m$^{-3}$ for the density of Fe XIV at the excited level of 530.3 nm \cite{Schmelz_2012, Habbal_2011}, which is about 40\% higher than the one we used in Ref.~\cite{Berera:2023fhd}.
Taking these factors into account, the rate of $\ket 2$-states per second is computed to be
$
N_2^{530.3}  \sim   2062,
$
while the rates of $\ket 3$-states and higher are still negligibly small.
Turning to the UV emission line of 19.664 nm, where both the intensity of the line \cite{Young_2008} and the density the excited Fe XII ions can acquire larger values, this leads to a rate of $\ket 2$-states $N_2^{19.664}\sim1957$.
Despite following practical guidelines, the rate of $\ket 2$-state for 1.5 nm X ray can not be materially improved.
The rates of $\ket 2$-states of optical and UV lines are listed in the table \ref{table3}.
\begin{table}[ht]
\begin{center}
\begin{tabular}{  c|c|c|c } 
$\lambda_0$ (nm) &  530.3  & 19.664 & 1.5  \\\hline
$N_1$ per s &  2.4$\times10^{13}$  &    1.56$\times10^{16}$ & 3.78$\times10^{10}$\\
$N_2$ per s &  2062  &    1957 & $\sim$ 0
\end{tabular}
\caption{\label{table3}
Estimates of the rates of $\ket 2$-states in the optical and UV range.}
\end{center}
\end{table}

\medskip

{\it Decoherence \& Discussion} -- Having shown the rates of production of these N-photon states, one may wonder how likely it is for such states to be received by a detector intact without decoherence.
Ref.~\cite{Berera:2023fhd} shows the decoherence of photons in the optical to X-ray range is negligible.
Therefore, here we mainly consider the possibility of decohering radio and microwave photons.
Since the energy of radio/microwave photons are low, they interact with charged particles in the corona predominantly through Thomson scattering, which has a cross section of $\sigma_{\rm Th} = 6.65 \times 10^{-29}\ {\rm m}^2$. Employing the electron density $n_e$ in the corona given by Allen and Baumbach \cite{wexler2019spacecraft}, and Edenhofer, et al. \cite{edenhofer1977time} respectively, we arrive to a MFP=$1/(n_e \sigma_{\rm Th})$ around $10^{17}$ m, which is 6 orders of magnitude longer than the distance between the Earth and the Sun.

Another decoherence process could be the scattering of photons off dust particles. The average dust flux measured by In-Situ Helios \cite{kruger2019interstellar} in heliocentric distance range from $0.3$ to $1.0$ AU is $(2.6 \pm 0.3) \times 10^{-6}\ {\rm m}^{-2}\ {\rm s}^{-1}$, with dust radius of $0.37\ \mu$m. The shortest wavelength of the radio-microwave photons in our discussion is 5 orders of magnitude longer than the radius of these dust particles. As a result, the scattering would be well described by classical wave diffraction, altering the phase of the photons collectively and leaving coherence unbroken. Dividing the flux by the measured impact speed $\sim$ 60 km/s leads to a number density of the dust of $4\times10^{-11}$ m$^{-3}$, which further suggests a MFP of $\sim 10^{24}$ m for the scattering. The long MFP for scattering off the dust means such interactions are not significant.

Faraday rotation may be a potential effect, where there is no direct particle interaction.
This effect leads to a polarization rotation of angle $\beta$ which depends on the wavelength $\lambda$,
the electron density $n_e(s)$ at point $s$ along the path from the solar surface $R_{\odot}$ to an observer $d$, and the component of the magnetic field $B_{\parallel}(s)$ in the direction of propagation. A magnetic field of size $B_{\parallel} \sim 100\ \mu$T is
an overestimate for the value of magnetic field for the entire region under consideration \cite{kooi2022modern}.
Using this value and following the calculation
in Ref.~\cite{Berera:2023fhd} for 
$\beta(\lambda=100\ {\rm nm})$,  yields the rotation angles for the relevant wavelengths in the radio-microwave range of
$$\beta(\lambda=100\ {\rm cm}) \simeq 10^{-4}, ~~\beta(\lambda=1\ {\rm cm}) \simeq 10^{-8}.$$
The electron density and magnetic field strength in Earth's atmosphere are of similar magnitudes as those in the solar corona, but Earth's much thinner atmosphere leads to even smaller rotation angle. This is a small
rotation and in addition Faraday rotation impacts photon in the aggregate. Thus, Faraday rotation results in merely an overall rotation without decoherence effects. 

The propagation of photons in the Earth's ionosphere could be subject to interactions because the varying electron density changes electric permittivity. However, the plasma frequency of the ionosphere is about 6 to 60 MHz\cite{Jackson:1998nia},  which is much lower than the frequencies considered here. As such, this would at most cause the electric permittivity to barely deviate  from 1 and only slightly refract the low frequency photons. Therefore, the ionosphere does not decohere photons
that have frequency much higher than 60 MHz.

Another effect, birefringence, usually splits a wave into different paths, depending on  the directions of  polarization and propagation. For example, $\ket{2_{\vec{k},s_1}}$ and $\ket{2_{\vec{k},s_2}}$ have polarization $s_1$ and $s_2$ respectively, and birefringence can split them but has no
relative effect on individual
components of $\ket{2_{\vec{k},s_1}}$ and $\ket{2_{\vec{k},s_2}}$.  This means 
for example it affects the two photons of $\ket{2_{\vec{k},s_1}}$ in the same manner, because they share the same polarization $s_1$.
Consequently, the occurrence of the birefringence does not imply coherence of the state is broken. Similar to the study of effects from gravity \cite{Berera:2022nzs}, it only induces a change in fidelity, which is distinguished from the loss of quantum coherence.

\medskip

{\it Measurements} -- Measuring the quantum nature of a signal in space could be
significantly more technically demanding than in a terrestrial lab.  
Since the interactions affecting photons in the radio-microwave frequency range either have very long MFPs or merely change  phase uniformly (e.g., We found negligible ionosphere effects and the small Faraday rotations including the Earth's atmosphere.), our investigations demonstrate that decoherence mechanisms  are ineffective against these photons. Therefore, a proposed test of photon coherence within the radio-microwave band can be Earth-based instead of relying on satellites, hence significantly reducing the difficulty and cost of conducting such experiments. Moreover the impact from the Earth's atmosphere can be further reduced by taking measurements on days without clouds\cite{Deng:2019der}.

Most of the indistinguishability tests of photons are based on the Hong-Ou-Mandel (HOM) effect\cite{Hong:1987zz}, 
in which
two identical single photons enter a beam splitter through different
input gates but leave through the same exit terminal.
Based on the HOM effect, Deng et al. \cite{Deng:2019der,duanadded} tested quantum interference in the optical range between single solar photons and photons produced by a quantum dot on the Earth, demonstrating distinct evidence for the quantum nature of light.
Within the frequency band of our investigation, HOM experiments for microwave photons ($\sim$ 7.25 GHz) from different sources have also been performed \cite{Lang2013}, which provides the blueprint for future experimental setups.

The conventional HOM effect is that two $\ket 1$ states coalesces into one $\ket 2$ state, while our focus is to test the coherence of a $\ket 2$ state from the start. Hence, to identify a $\ket 2$ state, a time reversed HOM setup should be used which suppresses two-photon components in each output port \cite{PhysRevLett.121.200502}, leading to one $\ket 1$-state coming out of each exit port.
The dimensional size of HOM test devices being around 1 mm\cite{Lang2013} sets the rate of $\ket 2$ around 10$^{0}$ to 10$^{1}$ s$^{-1}$ per HOM device.
Note that the areas involved in these estimates are presumed perpendicular to a radial trajectory from the Sun.

Let us begin with a rough description of the expected signal and noise from our proposed HOM tests of our rate estimations. Suppose the device has a working area of 1 mm$^2$, two entrances $a'$, $b'$ and two exits $a$, $b$. Picking 9.91 GHz as the example, a number of $N_1(9.91\text{ GHz})\times(10^{-3})^2$ $\ket1$ states would enter either $a'$ or $b'$ and come out of either $a$ or $b$, which should give background noise  with average time gap between adjacent $\ket1$ states being around 0.65 ms.
 
Overlapping with the $\ket1$ background are a number of $N_2(9.91\text{ GHz})\times(10^{-3})^2$ $\ket2$ states,
which also enter either $a'$ or $b'$ but come out of both $a$ and $b$ simultaneously with identical $\ket1$ pairs. This should give a signal with average time gap between adjacent $\ket1$ pairs being around 34.8 ms.   As such,
the average timing pattern in the $\ket2$ signals is substantially distinctive comparing to that of the background $\ket1$-photons. But note that the wavelength of these photons are about 3 cm, 
and so we expect them to be distinguishable at the level of $\times 0.1$ ns. Hence, both signal and noise rates are slow enough that we expect very few random coincidences.

Let us next investigate the ratio of signal to noise for photons received at random as listed in the tables in more detail.
Again we consider $\ket 1$ state photons of fixed frequency as noise and $\ket 2$ state photons of 
the same frequency as the signal. As above, we will focus on the 9.9 GHz case as an example. The
probability of no coincidence $P_{nc} $ in one second  in a 1 mm$^2$ HOM  device due to $\ket{1}$ state 
photons is
$$P_{nc} =  \frac{b!}{ (b-n)!  b^n}$$
where $b$ is the number of temporal bin (which we take to be $10^7$, i.e.,  the bins are 100 ns each)
and $n$ is the number of 9.9 GHz $\ket 1$ state photons per second on the device, which is 1540.
For $b$ large and $b\gg n$ the approximate  probability of a coincidence (i.e., fake signal) is 
$$P_c = 1- e^{^{\frac{-n(n-1)}{ 2b}}}=11\%.$$
The true resolution of the device is probably closer to 1 ns which would give $P_c=0.1\%$.
On the other hand, we assume all $\ket 2$ state photons convert to two $\ket 1$ state photons, one in each are of the HOM device, so all generate a coincidence. In the 9.9 GHz example that gives
a rate of 28 per s in a 1 mm$^2$ detector. Hence, the experiment is dominated by the signal in this example, and we arrive at the same conclusion for all frequencies listed in the tables. to summarize, if the inverse HOM effect is efficient, then multi-photon states can easily be detected from the solar corona.

We have also given the rates for high number states including $\ket 3$, $\ket 4$ etc. It is worth noting that three-photon interference has been observed \cite{PhysRevLett.118.153602} and there is developing research on the four-photon HOM effect \cite{PhysRevA.100.053829}.
We also note that other astrophysical quantum experiments in the optical region have been proposed by Dravins et al. \cite{2008ASSL..351...95D,dravins2005quanteye}.

\medskip

{\it Summary} -- Because of the relations between the production rate of  these quantum states and the properties of the solar corona, such as ionic content, thickness and temperature, experimental measurements of the quantum states should unveil new information on the structure of the corona.

Starting from the observational data of solar radio flux and the theory of stimulated emission, this work calculated the rates of potentially measurable N-identical photons states $\ket N$, originating from the propagation of photons through the stellar atmosphere, specifically, the solar corona.
We estimated the rate numbers of detectable $\ket N$-states of photons and gave the average expected results in future tests.
We demonstrated that various interactions including scattering, Faraday rotation, etc. are not effective in decohering these quantum states of photons, particularly due to the long MFPs between interactions.
We discussed how to measure these quantum states via the HOM effect and how the setups in the prior experiments could be modified and applied to the measurements. 

Quantum effect are quite prevalent in the fields of particle physics
and condensed matter physics though have been slow to emerge in
astronomy. This Letter has given an example of how
modern techniques developed in quantum optics laboratories
can be applied to probe the quantum nature of astronomy.
In particular, we have developed a method that uses quantum coherence to
provide a valuable alternative probe of the
solar atmospheric structure and possibly that of
nearby stars. At a fundamental level the measurement of these
stimulated emission states discussed in
this Letter would be a novel demonstration of
quantum coherence of a state sustained over astronomical distance.

\section*{Acknowledgments}

AB is partially funded by STFC. For the purpose of open access, the author has applied a Creative Commons Attribution (CC BY) licence to any Author Accepted Manuscript version arising from this submission. This work is supported in part by the National Key Research and Development Program of China under Grant No. 2020YFC2201501 and  the National Natural Science Foundation of China (NSFC) under Grant No. 12347103.

\bibliography{01}

\end{document}